\documentstyle[aps,preprint]{revtex} 
 
\tightenlines 
\begin{document}

\title{N-Site approximations and CAM analysis for a stochastic sandpile} 
 
\author{Ronald Dickman$\dagger$} 
\address{ 
Departamento de F\'{\i} sica, ICEx, 
Universidade Federal de Minas Gerais, 
Caixa Postal 702, 
30161-970 Belo Horizonte, MG, Brazil\\ 
} 
\date{\today} 
 
\maketitle 
\begin{abstract} 
I develop n-site cluster approximations for a stochastic  
sandpile in one dimension.  A height restriction is imposed
to limit the number of states: 
each site can harbor at most two particles (height $z_i \leq 2$). 
(This yields a considerable simplification  
over the unrestricted case, in which the number of states 
per site is unbounded.)   
On the basis of results for $n \leq 11$ sites, I estimate the 
critical particle density as $\zeta_c = 0.930(1)$, in good agreement 
with simulations.  A coherent anomaly analysis yields 
estimates for  
the order parameter exponent [$\beta = 0.41(1)$] and the 
relaxation time exponent ($\nu_{||} \simeq 2.5$). 
\end{abstract} 
 
\pacs{PACS numbers: 05.40.+j, 05.70.Ln } 
 
 
\date{\today} 
 
\section{Introduction} 
 
Sandpile models are the prime example of self-organized  
criticality (SOC) \cite{btw,dhar99}, in which 
a system with an absorbing-state phase 
transition is forced to its critical point \cite{bjp,granada,cancun}, 
leading to scale-invariance in the apparent absence of 
parameters \cite{ggrin}.  The absorbing-state phase transition, 
which depends, as is usual, on the fine-tuning of one or more 
control parameters, is evident in sandpiles with a fixed number of  
particles \cite{bjp,tb88,vz,dvz,vdmz}, models that have 
come to be called fixed-energy sandpiles (FES).   
While most studies of sandpiles have probed 
the driven case \cite{dhar99,priez}, there is great interest in understanding  
the scaling properties of FES models as well \cite{dvz,mdvz,chessa,carlson}.   
 
Previous studies of FES reveal that they 
exhibit a phase transition between an absorbing and an active state 
as the particle density $\zeta$ (which is the temperaturelike control parameter) 
is increased beyond a critical value 
\cite{bjp,fes2,m1d}.  Until now, all quantitative results on FES 
have been obtained from simulations.  It is therefore of interest 
to apply theoretical methods to such models.  One such approach 
is Suzuki's coherent anomaly method (CAM) for analyzing a series of 
cluster approximations.  It has been shown to yield good estimates 
for critical properties both in \cite{suzuki86,suzuki87} and out 
of equilibrium \cite{konno,ferreira}.  In this work I develop 
$n$-site approximations for a one-dimensional  
sandpile model, and analyze the results using the CAM. 
This represents the first application of the CAM to a model 
representative of the class of 
absorbing-state phase transitions in systems with a conserved 
density \cite{ale00,pastor,lubeck}. 
 
In this paper I study a FES with a height restriction. 
From the theoretical viewpoint, an inconvenient feature of sandpile  
models is the unbounded number of particles that may occupy the
same site; this  
complicates attempts to derive cluster approximations.  In Manna's 
stochastic sandpile \cite{manna,manna2}, sites with height $z \geq 2$ are active. 
If we restrict the height (or number of sand grains per site), 
to be $\leq 2$, the effect on critical properties should be minimal, 
aside from a possible shift in the critical density $\zeta_c$. 
This expectation was recently verified numerically: the restricted-height 
stochastic sandpile belongs to the same universality class as its 
unrestricted counterpart \cite{mnrst}.  I study the restricted-height 
model for calculational convenience, as a representative of a broader 
universality class that includes Manna's stochastic sandpile, 
the conserved lattice gas, and the conserved threshold transfer 
process \cite{ale00}.

The balance of the paper is organized 
as follows.  The model is defined in
Sec. II, followed by a discussion of $n$-site 
approximations in Sec. III.   
Numerical results are presented in Sec. IV. The CAM analysis is discussed in
Sec. V, and in Sec. VI I present a brief summary.
 
\section{Model} 
 
The model is defined on a ring of $L$ sites 
with periodic boundaries.  (The cluster approximations 
effectively study the $L \to \infty$ limit.)  The 
configuration is specified by the number of particles 
$z_i = 0, 1$, or 2 at each site; sites with $z_i \!=\! 2$ are said to be 
{\it active}, and have a toppling rate of unity.  The  
continuous-time (sequential), Markovian dynamics  
consists of a series of toppling events at individual sites. 
When site $i$ topples, two particles attempt move to randomly  
chosen nearest neighbors $j$ and $j'$ of $i$.  The new position of 
each particle is accepted if and only if the target site has fewer than
two particles.  I consider a stochastic toppling rule in which  
the two particles move independently. 
Any particle attempting to move to a site already harboring two particles is 
sent back to the toppling site.  (Thus an attempt to send two particles 
from site $j$ to site $k$, with $z_k = 1$, results in $z_k\!=\!2$ 
and $z_j=1$.)  Transition probabilities are listed in Table I. 

This model, and a closely related one 
(with a cooperative toppling rule), were studied via simulation
in Ref. \cite{mnrst}, which showed that the critical exponents $\beta$ 
and $\nu_\perp$ are the same as for the unrestricted Manna sandpile.
A similar conclusion was reached in Ref. \cite{ale00} for a two-dimensional
restricted-height sandpile (called the conserved threshold transfer process
in that work).

\section{Cluster approximations} 
 
I have derived dynamic $n$-site cluster approximations for the 
one-dimensional restricted sandpile model.  Such approximations often 
yield qualitatively correct phase diagrams \cite{marro}. 
The procedure parallels that used by Ferreira and Mendirata  
to study the one-dimensional contact process \cite{ferreira}. 
The $n$-site approximation consists of a set of coupled differential  
equations for the probabilities $P_{\cal C}^{(n)}$ of each $n$-site  
configuration, ${\cal C}$. 
(There are $3^n$ such configurations, but the number of independent 
probabilities is $\sim 3^n/2$, due to symmetries.) 
The system is assumed homogeneous, so that the $P_{\cal C}^{(n)}$  
are independent of position.   
 
Since transitions in a set of $n$ contiguous sites generally 
depend on sites outside the cluster, the $n$-site probabilities 
are coupled to those for $n\!+\!1$ and so on, generating an infinite 
hierarchy of equations.  The $n$-site approximation truncates this 
hierarchy by approximating $m$-site probabilities (for $m>n$) 
in terms of $n$-site {\it conditional} probabilitites. 
In the $n$-site approximation, the joint probability 
for a sequence of $n\!+\!1$ sites is approximated so \cite{ferreira}: 
 
\begin{eqnarray} 
\nonumber
P^{(n+1)} (z_1,...,z_{n+1}) & \simeq &  
P^{(n)} (z_{n+1}|z_n,...,z_2) P^{(n)} (z_n,...,z_1)     
\\
&=& \frac{P^{(n)} (z_{n+1},...,z_2) P^{(n)} (z_n,...,z_1)} 
{P^{(n-1)}(z_n,...,z_2)} \;. 
\label{factpa} 
\end{eqnarray} 
 
The equations for one- and two-site approximations are relatively simple to  
derive, and are described in Ref. \cite{mnrst}. 
I have developed a computational algorithm capable of generating the
approximation for arbitrary $n$.   
Each configuration ${\cal C} = (z_n,...,z_1)$ is represented by an 
integer  
 
\begin{equation} 
I({\cal C}) = \sum_{k=1}^n z_k \cdot 3^{k-1} \;. 
\end{equation} 
 
The calculation begins with the generation of all configurations, 
corresponding to each integer from zero (all sites empty) 
up to the maximum, $3^n-1$ (all sites doubly occupied); the 
symmetry (under inversion) of each configuration is determined. 
If  ${\cal C} $ is not symmetric, then it and its mirror image  
${\cal C}_R $ must have the same probability, and   
only the smaller of ${\cal C} $ and ${\cal C}_R $ is treated explicitly, 
reducing the number of variables by roughly half. 
 
Next, a list of all possible transitions is constructed.  Here it is 
useful to distinguish between {\it central} transitions (involving a  
toppling at one of the sites 2,...,$n\!-\!1$) and  
{\it boundary} transitions, in which either site 1 or site $n$, 
or one of the peripheral sites (0 or $n\!+\!1$) topples.   
The rate of a central transition ${\cal C}  \to {\cal C}' $ 
is the product of a branching probability $p_b$  
(for the particles to be redistributed in a particular manner, as
in Table I),  
and the intrinsic toppling rate, which is unity.  Consider, for 
example, configuration ${\cal C} = (21120)$.  The transition rates 
associated with a toppling at the second site 
(counting from the right), are 
\[ 
w[(21120) \to (21102)] = \frac{1}{4} ,
\] 
\[ 
w[(21120) \to (21201)] = \frac{1}{2} ,
\] 
\[ 
w[(21120) \to (21210)] = \frac{1}{4} .
\] 
 
For each configuration ${\cal C}$, the set of allowed transitions 
to other states, and the associated rates, are stored.  
In the case of a central transition from ${\cal C}$ to
${\cal C'}$, the contribution to the 
time-derivative of the probability has the form: 
\[ 
\frac{d P({\cal C'})}{dt} = w[{\cal C} \to {\cal C}'] P({\cal C}). 
\] 
There is of course a corresponding loss term for $P({\cal C})$: 
\[ 
\frac{d P({\cal C})}{dt} = -w[{\cal C} \to {\cal C}'] P({\cal C}). 
\] 
Thus central transitions contribute to the evolution of the 
probability distribution precisely as in the 
master equation.   
 
For boundary transitions, one does not 
have access to the $n\!+\!1$-site 
probabilitites required to mount a complete description, 
and so must resort to the truncation scheme embodied in Eq. (\ref{factpa}). 
For example, the contribution to $dP(2,z_2,...,z_n)/dt$  due to 
the transition $(2,z_2,...,z_n) \to (1,z_2,...,z_n )$ is 
\[ 
\frac{1}{2} P^{(n+1)}(0,2,z_2,...,z_n) +  
\frac{3}{4} P^{(n+1)}(1,2,z_2,...,z_n) ~.
\] 
The $P^{(n+1)}$ are estimated using Eq. (\ref{factpa}).
For boundary transitions one stores not only the 
rate, but the two configurations (aside from the original one, 
${\cal C}$) whose probabilities are needed to evaluate  
$dP({\cal C})/dt$.  With this information available,  
one can evaluate the derivatives 
$dP({\cal C})/dt$ for all possible configurations, given the 
probability distribution.    
 
The evolution of the probability distribution is found via numerical 
integration, using a fourth-order Runge-Kutta scheme \cite{numrec}. 
The integration is halted when a stationary distribution is 
attained, that is, when the time-derivatives $dP({\cal C})/dt$ all 
have an absolute value smaller than $\delta$ (typically,
$\delta = 10^{-13}$).   
An interesting technical point concerns the evaluation 
of the $n\!-\!1$-site marginal distribution.  There are evidently  
two equivalent expressions that may be used: 
\[ 
P^{(n-1)}(z_{n-1},...,z_{1}) = \sum_{z_n = 0}^2 
P^{(n)}(z_n,z_{n-1},...,z_{1})  \;, 
\] 
and 
\[ 
P^{(n-1)}(z_{n-1},...,z_{1}) = \sum_{z_0 = 0}^2 
P^{(n)}(z_{n-1},...,z_{1},z_0)  \;. 
\] 
Numerical stability is {\it greatly enhanced} using the 
{\it mean} of the two expressions given above. 
 
For  
sizes $n \geq 7$, very near the critical point, relaxation to the 
stationary distribution is very slow, and the following procedure 
proves advantageous.  Let $D = \max_{\cal C} |dP({\cal C})/dt|$ be 
the largest derivative (in absolute value). 
The properties of interest (principally, 
the active-site density) are recorded as a function of $D$, and the  
integration halted when $D < 10^{-10}$.  Fig. 1 shows the result for the 
active site density, for $\zeta $ slightly above $\zeta_c$. 
The stationary value is obtained via extrapolation to $D = 0$, 
usually via a quadratic fit to the four data points for smallest $D$. 
(The resulting correction is typically less than 1\% of the value 
at $D = 10^{-10}$.) 
I also studied the order parameter  
relaxation rate $\gamma = |\dot{\rho}/\rho| $ for each $n$ 
at a series of $\zeta$ values near, but below, $\zeta_c$.  
These data are used to estimate the critical exponent $\nu_{||}$ 
in Sec. V.

\section{Numerical results} 
 
I derived cluster approximations for $n \leq 11$ sites, 
yielding the stationary active-site density $\rho_a$ as a function of $\zeta$. 
The $n$-site approximation predictions for $\rho_a(\zeta)$ 
($n \!=\! 3$ to 11) 
are compared against simulation in Fig. 2; the theoretical 
curves appear to approach the simulation result systematically. 
 
For each $n$, the active-site density is zero below a certain 
critical value, $\zeta_{c,n}$.  Since the phase transition in the 
stochastic sandpile is continuous \cite{m1d,mnrst},  one 
expects the same to be true of the cluster approximations. 
This is indeed the case for $n \leq 4$, but for $n = 5$, 6 and 7 there 
is a very small discontinuity in $\rho_a$ ($\stackrel < \sim  10^{-3}$, 
invisible on the scale of Fig. 2), as we 
decrease $\zeta$.  Since the same procedure is used for 
all $n$, the discontinuity is unlikely to be artefact of 
the numerical method.  On the other hand, I do not 
regard the discontinuity as physically significant; it appears to  
represent an unphysical feature of the cluster approximations for 
certain $n$ values, very near the critical 
density.  In the CAM analysis I disregard the behavior of $\rho_a$ 
in the immediate vicinity of $\zeta_{c,n}$, and instead  
analyze its properties at points somewhat removed from the transition.

There remains, naturally, the problem of estimating the critical 
density, $\zeta_{c,n}$.  For each $n$ I determine the critical density 
by fitting the four or five data points nearest the transition, where 
$\rho \simeq 10^{-3}$ or less.  In each case, I plot $\ln \rho_a$ 
versus $\ln (\zeta - \zeta_{c,n})$, varying $\zeta_{c,n}$ to obtain 
the best power-law fit.  The associated slopes vary between 
1 (for $n\!=\!3$) and about 0.25 (for $n\!=\!6$), but these, again,  
are regarded as unimportant details of the approximation 
in question.  It is important to stress that, in the cases where the 
transition is apparently discontinuous, the difference between the 
location of the discontinuity and the extrapolated value of 
$\zeta_{c,n}$ is less than one part in 10$^5$, and that the estimates 
for $\zeta_c$ and critical exponents are insensitive to these 
tiny differences. 
The values of $\zeta_{c,n}$ obtained in this manner are listed in Table II. 
 
Using the results for $\zeta_{c,n}$, I estimate  
$\zeta_c = \lim_{n \to \infty} \zeta_{c,n}$ by plotting 
$\Delta_n \equiv \zeta_c - \zeta_{c,n} $ versus $ n$ in a double-logarithmic
plot, varying $\zeta_c$ 
to obtain the best power-law fit.  The latter is obtained using $\zeta_c$ 
in the range 0.929 - 0.931, yielding $\zeta_c = 0.930(1)$, in good 
agreement with the simulation result of 0.92965 \cite{mnrst}. 
The finite-size scaling prediction 
for the critical point shift 
is \cite{fss}: $\Delta_n^{\nu_\perp} \propto 
1/n$.  I obtain a good fit to the data (see Fig. 3) using $\nu_\perp = 1.66$  
(as found in simulations \cite{m1d,mnrst}), including 
a correction to scaling term: 
\[ 
\Delta_n^{\nu_\perp} \propto \frac{A}{n} + \frac{B}{n^2} \;. 
\] 
The numerical data are consistent with the 
simulation estimate for $\nu_\perp$, but not sufficient
to furnish an independent estimate of the exponent. 
 
\section{Coherent anomaly analysis} 
 
A detailed explanation of the CAM procedure is given in  
Ref. \cite{suzuki86,suzuki87}; it may be understood 
on the basis of finite-size scaling \cite{fss}.  The
approach here parallels that used by Tom\'e and de Oliveira
in their study of the Domany-Kinzel model \cite{tomecam}.
To begin, one argues that the cluster size $n$ plays the 
role of an effective system size $L$ as regards scaling properties. 
This is because the $n$-site approximation effectively cuts off 
correlations of range $> n$ (notwithstanding the fact that 
cluster approximations nominally treat an infinite system).   
Thus, as noted 
above, one expects a critical point shift $\Delta_n 
\propto n^{-1/\nu_\perp}$.  Finite-size scaling theory also 
yields the relation $\rho_{a,n} (\zeta_c) \propto n^{-\beta/\nu_\perp} 
\propto \Delta_n^\beta$ 
for the order parameter in a finite system, at the (true) critical 
point.  For $\zeta > \zeta_{c,n}$, 
$\rho_{a,n} (\zeta)$ is a smooth function.  Thus we are led to a 
scaling hypothesis for the order parameter \cite{tomecam}: 
 
\begin{equation} 
\rho_{a,n} (\zeta) =  
\Delta_n^\beta \mbox{\large $f$}  
\left(\! \frac{\zeta\!-\!\zeta_{c,n}} {\Delta_n} \! \right), 
\label{schyp} 
\end{equation} 
where $f(x)$ is a scaling function with $f(0) \!=\!0$. 
If we suppose that $f(x) \propto x^{\beta_{MF}}$ for $0 \leq x \leq 1$, 
then $\rho_{a,n} (\zeta) = A_n (\zeta - \zeta_{c,n})^{\beta_{MF}}$, 
where the amplitude 
$A_n$ diverges as $n \to \infty$:
 
\begin{equation} 
A_n \propto \Delta_n^{-(\beta_{MF}-\beta)} . 
\label{ampdiv} 
\end{equation} 
This is the usual CAM relation.  On the other hand, the hypothesis that 
$n$ is equivalent to a finite system size leads directly to: 
 
\begin{equation} 
\rho_{a,n} (\zeta_c) \sim \Delta_n^\beta . 
\label{scrhc} 
\end{equation} 
This expression involves the behavior of the 
$n$-site approximation at the critical point $\zeta_c$ not $\zeta_{c,n}$. 
It is interesting to note that the hypothesis 
of an effective system size directly proportional to $n$ is not strictly  
necessary.  The scaling relations involving $\Delta_n$ follow from  
the more general hypothesis of an effective system size  
$L_{eff} = L_{eff}(n)$, for example $L_{eff} \propto n^\phi$ with 
$\phi > 0$.

\subsection{CAM analysis for $\beta$} 
 
As noted above, the $n$-site approximations for the order parameter 
$\rho_{a,n}$ are not all well behaved in the vicinity of $\zeta_{c,n}$. 
For this reason, analysis of $\rho_{a,n}$ at $\zeta_{c,n}$ will not 
yield a consistent set of well defined amplitudes $A_n$. 
But since $\rho_{a,n} (\zeta) $ is well behaved for $\zeta > \zeta_{c,n}$, 
we can study its scaling at some point intermediate between $\zeta_{c,n}$ 
and $\zeta_c$.  In particular, the scaling hypothesis Eq. (\ref{schyp}) 
implies that if we fix $x = (\zeta\!-\!\zeta_{c,n})/\Delta_n $, 
then 
 
\begin{equation} 
\frac{d\rho_{a,n}}{d\zeta} = f(x) \Delta_n^{\beta-\beta_{MF}} . 
\label{scder} 
\end{equation} 
Our strategy is to analyze the order parameter data reasonably near the 
$n$-site critical value, but away for $\zeta_{c,n}$ itself, where 
$\rho_{a,n}$ is singular.  A crucial point in this analysis is the 
postulate that the mean-field exponent $\beta_{MF} \!=\! 1$, 
{\it regardless of the behavior of} $\rho_{a,n}$ {\it in the immediate 
vicinity of} $\zeta_{c,n}$.  The motivation for this assumption is,  
firstly, that $\beta_{MF}$ is clearly unity for $n \!=\!1$, 2, or 3; 
secondly, that a critical exponent such as $\beta$ is determined, 
in mean-field theory, by symmetry properties of the order parameter, 
and hence should not vary with $n$; and thirdly, that $\beta_{MF} \!=\!1$ 
generically for phase transitions to an absorbing state \cite{marro}. 
(The basis for this last assertion is that 
the mean-field equation for the order parameter will have the
form $d\rho/dt = A\rho - B\rho^2$, barring
some coincidence or a symmetry that renders $A$ and/or $B$ zero 
\cite{glb}.)
 
I evaluate $d \rho_{a,n}/d\zeta$ (numerically, using an interval 
$\Delta \!=\! 0.0005$), for fixed $x\!=\!1/4$; the results are 
shown in Fig. 4.  Least-squares linear fits to the data for 
$n \!=\! 8$ - 11 yield, via Eq. (\ref{scder}), 
the value $\beta = 0.408(6)$, where the figure in parenthesis denotes the 
uncertainty.  Using $\zeta_c \!=\! 0.929$ instead of the best estimate, 
0.930, I find $\beta = 0.421(5)$.  Thus a reasonable 
estimate for $\beta$ is 0.41(1).  (A similar analysis, but evaluating 
the derivatives at $x \!= \! 1/2$, yields $\beta = 0.42$.) 
 
The above analysis is complemented with a study of $\rho_{a,n} (\zeta_c)$, 
as suggested by Eq. (\ref{scrhc}).  The graph of $\rho_{a,n} (\zeta_c)$ 
versus $\Delta_n$ shows (on log scales, see Fig. 5), a fair amount 
of curvature, making determination of $\beta$ more difficult in this 
case.  Linear fits to the data for $n$ = 7-9, 8-10, and 9-11 yield, 
respectively, $\beta$ = 0.471, 0.460, and 0.448, consistent with 
an approach to the value of $0.41$ for large $n$.  Verification of 
convergence must naturally await the evaluation of  
approximations for larger clusters. 
 
A further point of interest is the validity of the scaling hypothesis, 
Eq. (\ref{schyp}).  The data collapse shown in Fig. 6, a plot  
of $\rho^* = \Delta_n^{-\beta} \rho_{a,n} $ versus  
$x = (\zeta\!-\!\zeta_{c,n})/\Delta_n $ 
provides support for the hypothesis. 
(In Ref. \cite{tomecam} a similar collapse is demonstrated 
for the Domany-Kinzel model.)
 
\subsection{CAM analysis for $\nu_{||}$} 
 
As shown in Refs. \cite{suzuki86,suzuki87,konno}, the CAM approach is 
readily extended to dynamics.  Let  
\begin{equation} 
\gamma_n(\zeta) = -\frac{1}{\rho}\frac{d\rho}{dt} 
\label{gamman} 
\end{equation} 
be the relaxation rate in the $n$-site approximation, and let
$\gamma (\zeta)$ be the true relaxation rate.  Then we expect 
$\gamma \sim |\zeta - \zeta_c|^{\nu_{||}}$, while  
$\gamma_n \sim |\zeta - \zeta_{c,n}|^{\nu_{||,MF}}$ in the $n$-site 
approximation, where the mean-field exponent is $\nu_{||,MF} \!=\! 1$ 
for models with an absorbing-state phase transition \cite{marro}. 
 A scaling hypothesis, analogous to Eq. (\ref{schyp}), for the 
relaxation rate, is 
\begin{equation} 
\gamma_{n} (\zeta) =  
\Delta_n^{\nu_{||}} \mbox{\large $g$}  
\left(\! \frac{\zeta\!-\!\zeta_{c,n}} {\Delta_n} \! \right), 
\label{schyrt} 
\end{equation} 
where the scaling function $g$ vanishes when its argument is 
zero.  Supposing that $g(x) \sim |x|^{\nu_{||,MF}}$, 
we see that  
$\gamma_{n} (\zeta) \sim \overline{\gamma}_n |\zeta\!-\!\zeta_{c,n}|$, 
where the amplitude follows 
\begin{equation} 
\overline{\gamma}_n \sim \Delta_n^{\nu_{||}-\nu_{||,MF}} . 
\label{camgam} 
\end{equation} 
 
I determine the relaxation rate numerically for $\zeta \stackrel < \sim  
\zeta_{c,n}$, and from these data extract the amplitudes  
$\overline{\gamma}_n$.  The results, shown in Fig. 7, display 
substantial curvature on a log-log plot, so that direct determination 
of the critical exponent $\nu_{||}$ is not feasible.  Simulations 
\cite{m1d,mnrst} yield estimates for $\nu_{||}$ in the range 
2.3 - 2.6.  The CAM results are consistent with values in 
this range, if we include a correction to scaling.  The solid line 
in Fig. 7 is given by 
\begin{equation} 
\ln \overline{\gamma}_n = 1.5 \ln \Delta_n + A \Delta_n - B, 
\end{equation} 
with fit parameters $A \!=\! 7.435$ and $B \!=\! 2.125$, consistent 
with a correction to scaling expression 
$\overline{\gamma}_n \propto \Delta_n^{\nu_{||}-1}(1 \!+\!A \Delta_n)$, 
with $\nu_{||} = 2.5$.  While results for larger clusters will be 
needed to determine $\nu_{||}$ with precision, one can at least assert 
that the present results are consistent with the rather imprecise 
estimates from simulations. 
 
\section{Discussion} 
 
I have devised a computational algorithm for generating $n$-site 
cluster approximations for a one-dimensional stochastic sandpile 
model with a fixed particle density.  To facilitate the analysis I 
impose the height restriction $z_i \leq 2$.  Analyzing the results for 
$n \leq 11$, I obtain the estimates $\zeta_c = 0.930(1)$, 
$\beta = 0.41(1)$ and $\nu_{||} \simeq 2.5$, all in agreement 
with simulation \cite{mnrst}.  While the results are not very precise, 
they provide significant independent support for the simulational 
findings, showing that FES models with a strict activity threshold 
belong to a universality class distinct from that of directed percolation 
(DP) \cite{ale00,pastor,lubeck}. 
Sandpile models in which sites with an above-threshold height can remain 
stable (so-called ``sticky grains"), have recently been shown to belong to the 
DP class \cite{mohanty}, but such is not the case for the model studied here. 
Application of the methods used in this work to other 
sandpile models should prove illuminating. 
\vspace{2em}

\noindent{\bf Acknowledgments} 
\vspace{1em} 

I thank M\'ario de Oliveira, T\^ania Tom\'e, Attila Szolnoki, Gyorgy Szab\'o,
G\'eza \'Odor, and Paulo Alfredo Gon\c{c}alves Penido for helpful 
comments.  This work was supported by CNPq, and CAPES, Brazil. 
\vspace{5em}

\noindent$^\dagger$ e-mail: dickman@fisica.ufmg.br

\newpage 
\begin{table} 
\begin{center} 
\begin{tabular}{|c|c|} 
Transition         &  Probability   \\         
\hline 
$020 \to 101$      &    1/2         \\  
$\;\;\;\; \to 200$ &    1/4        \\  
\hline   
$120 \to 201$      &    1/2        \\  
$\;\;\;\; \to 102$ &    1/4         \\  
$\;\;\;\; \to 210$ &    1/4         \\  
\hline   
$121 \to 202$      &    1/2          \\ 
$\;\;\;\; \to 211$ &    1/4       \\ 
\hline   
$220 \to 202$      &    1/4          \\ 
$\;\;\;\; \to 211$ &    1/2       \\ 
\hline 
$122 \to 212$      &    3/4        \\  
\end{tabular} 
\end{center} 
\label{tab1} 
\noindent{Transition probabilities for the restricted-height 
sandpile.   Probabilitites are symmetric under reflection.}
\end{table} 
 
\begin{table} 
\begin{center} 
\begin{tabular}{|c|c|} 
$n$  &  $\zeta_{c,n}$   \\         
\hline 
\hline 
1   & 0.5  \\  
2  & 0.75   \\  
3  & 0.80854  \\  
4 & 0.83682 \\  
5 & 0.85305  \\  
6   & 0.86378  \\  
7  & 0.87148   \\  
8  & 0.87736 \\  
9 & 0.88207 \\  
10 & 0.88594  \\  
11 &  0.88918            \\ 
\end{tabular} 
\end{center} 
\label{tab3} 
\noindent{Table II. Critical densities in the $n$-site approximation. 
}   
\end{table}

\newpage 
\noindent FIGURE CAPTIONS 
\vspace{1em} 
 
\noindent FIG. 1.  
Active-site density $\rho$ versus 
$D = \max_{\cal C} ~|dP({\cal C})/dt|$, for $n\!=\!10$, 
$\zeta \!=\! 0.8860$. 
\vspace{1em} 
 
\noindent FIG. 2.  
Stationary active-site density $\rho$ versus particle density $\zeta$. 
Solid curves: $n$-site approximations for $n$ = 3 - 11; 
points: simulation results for a system of 5000 sites. 
\vspace{1em} 
 
\noindent FIG. 3.  Critical point shift  
$\Delta_n^{\nu_\perp}$ versus$ 1/n$.   
The solid line is a fit including a correction term as described in the text. 
\vspace{1em} 
 
\noindent FIG. 4.  
$\rho_n' \equiv d\rho_{a,n}/d\zeta $ versus $\Delta_n$.   
The derivative is evaluated at $x\!=\!1/4$.  The slope of the solid 
line is -0.592, corresponding to $\beta = 0.408$. 
\vspace{1em} 
 
\noindent FIG. 5.  Active-site density
$\rho_{a,n}(\zeta_c) $ versus $\delta_n$.  The slope of the solid line is 0.408.  
\vspace{1em} 
 
\noindent FIG. 6. Scaled active-site density
$\rho^* = \Delta_n^{-\beta} \rho_{a,n} $ versus scaled particle density 
$x = (\zeta\!-\!\zeta_{c,n})/\Delta_n $, for $n\!=\!6$ - 11. 
\vspace{1em} 
 
\noindent FIG. 7. Relaxation rate amplitude $\overline{\gamma}_n$ 
versus $\Delta_n$.  The solid curve is a fit including 
a correction to scaling term as described in the text.
\vspace{1em} 
 
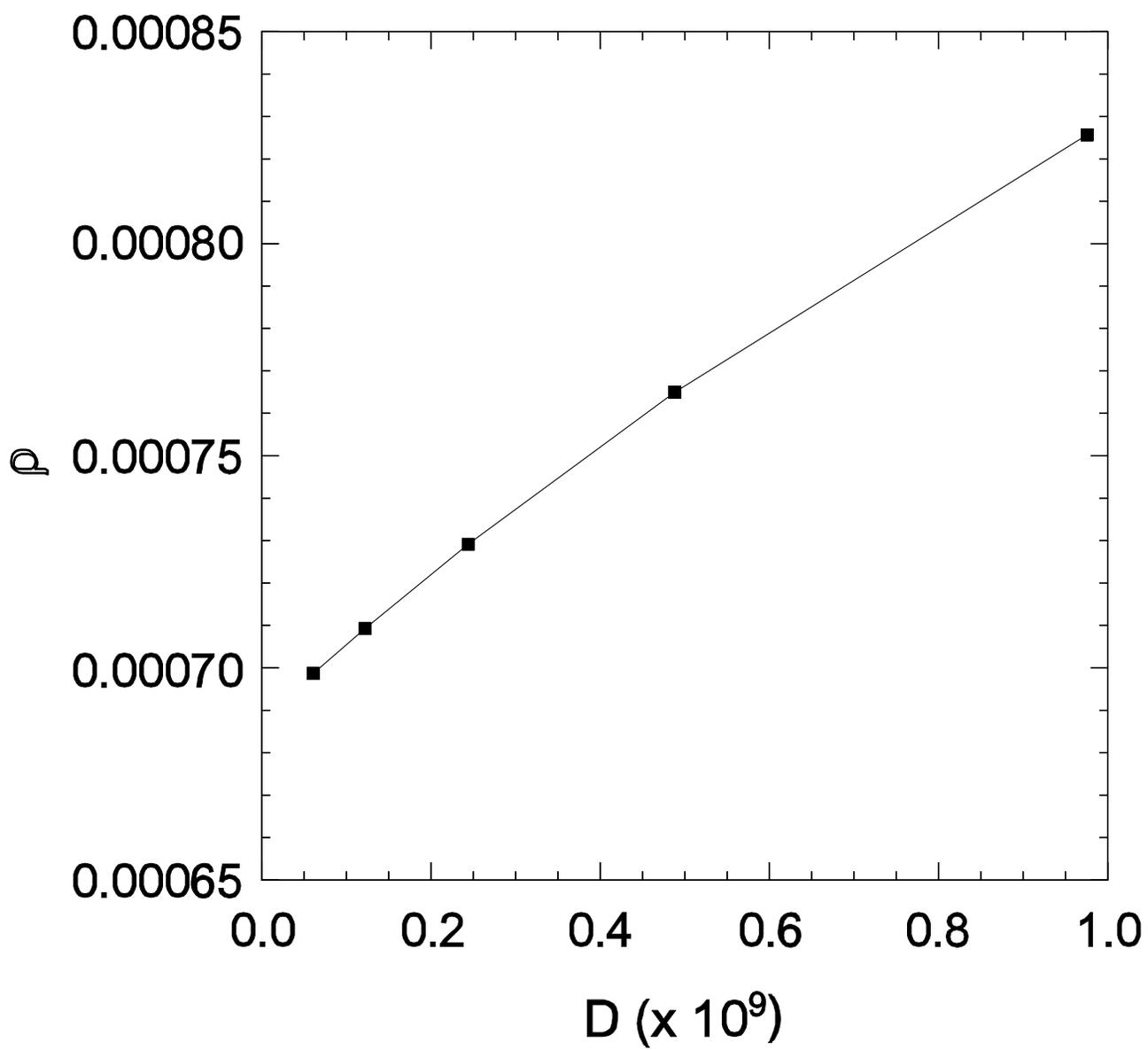
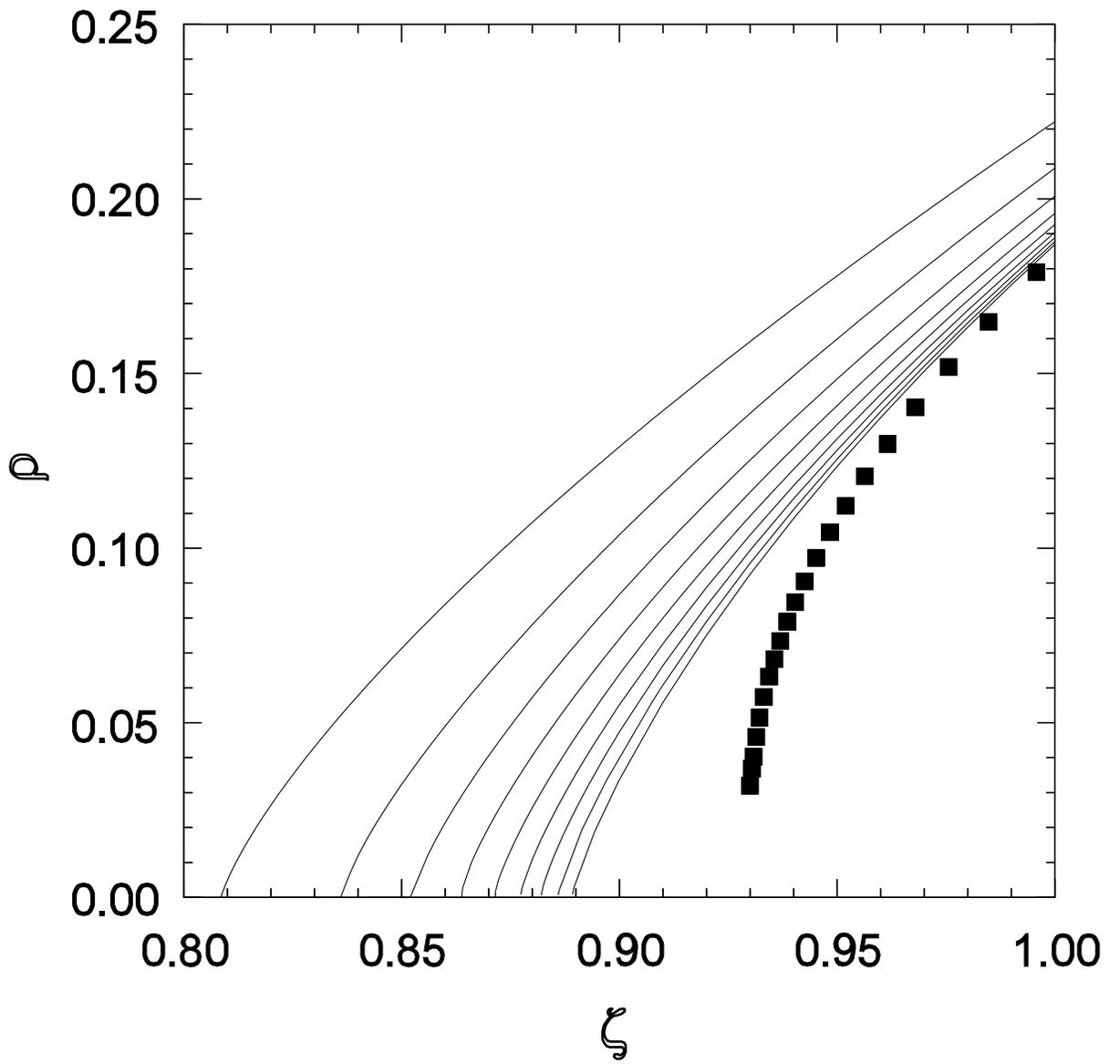
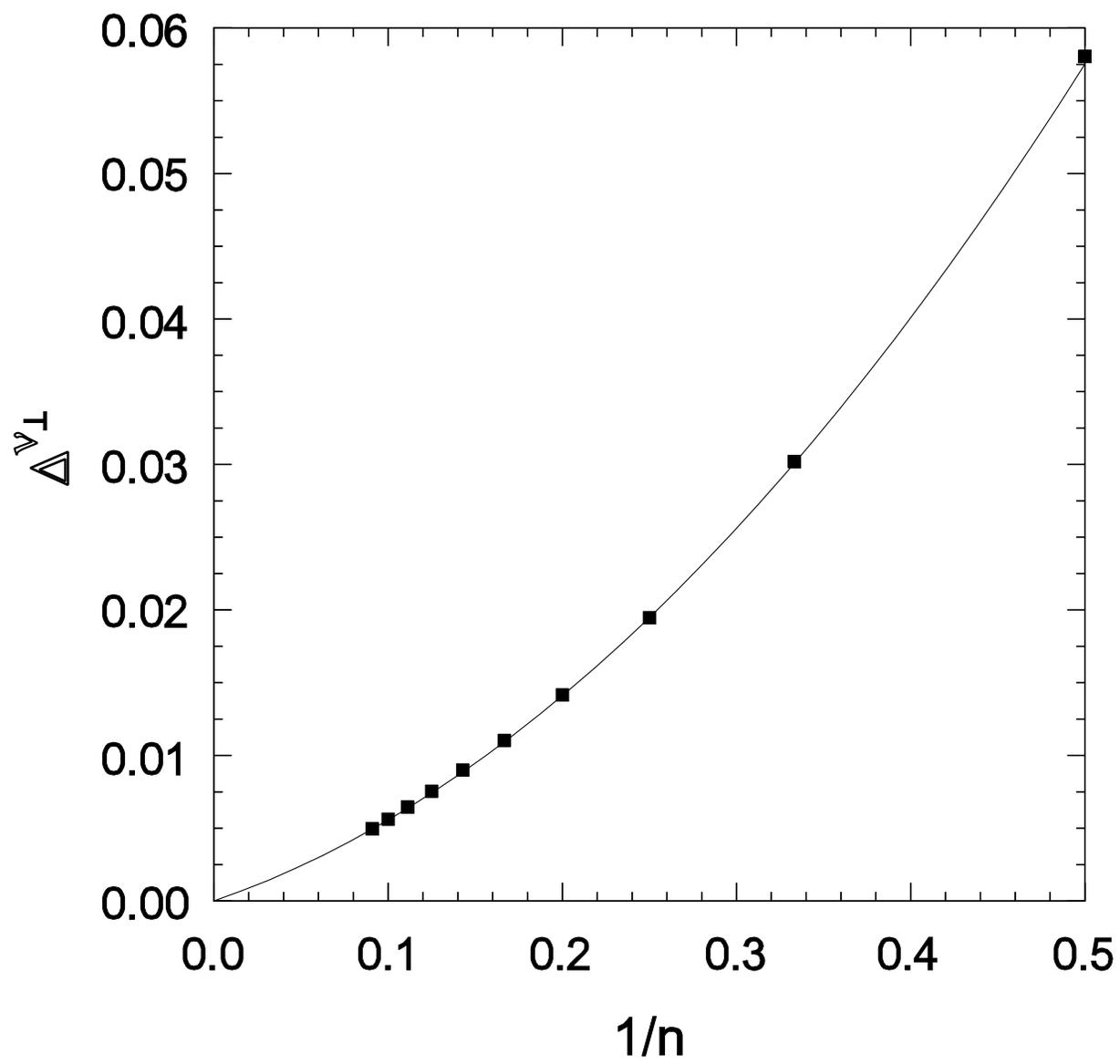
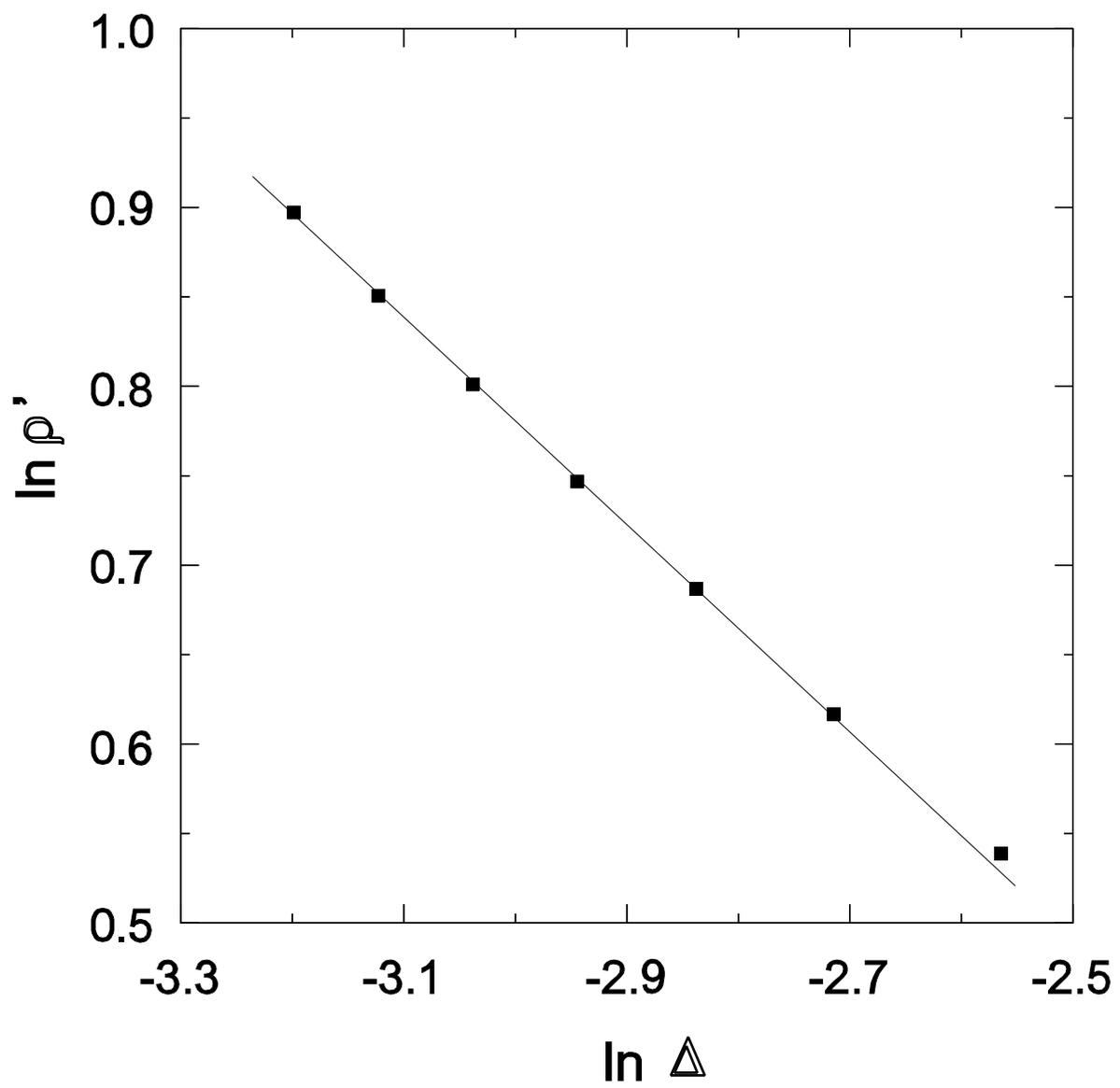
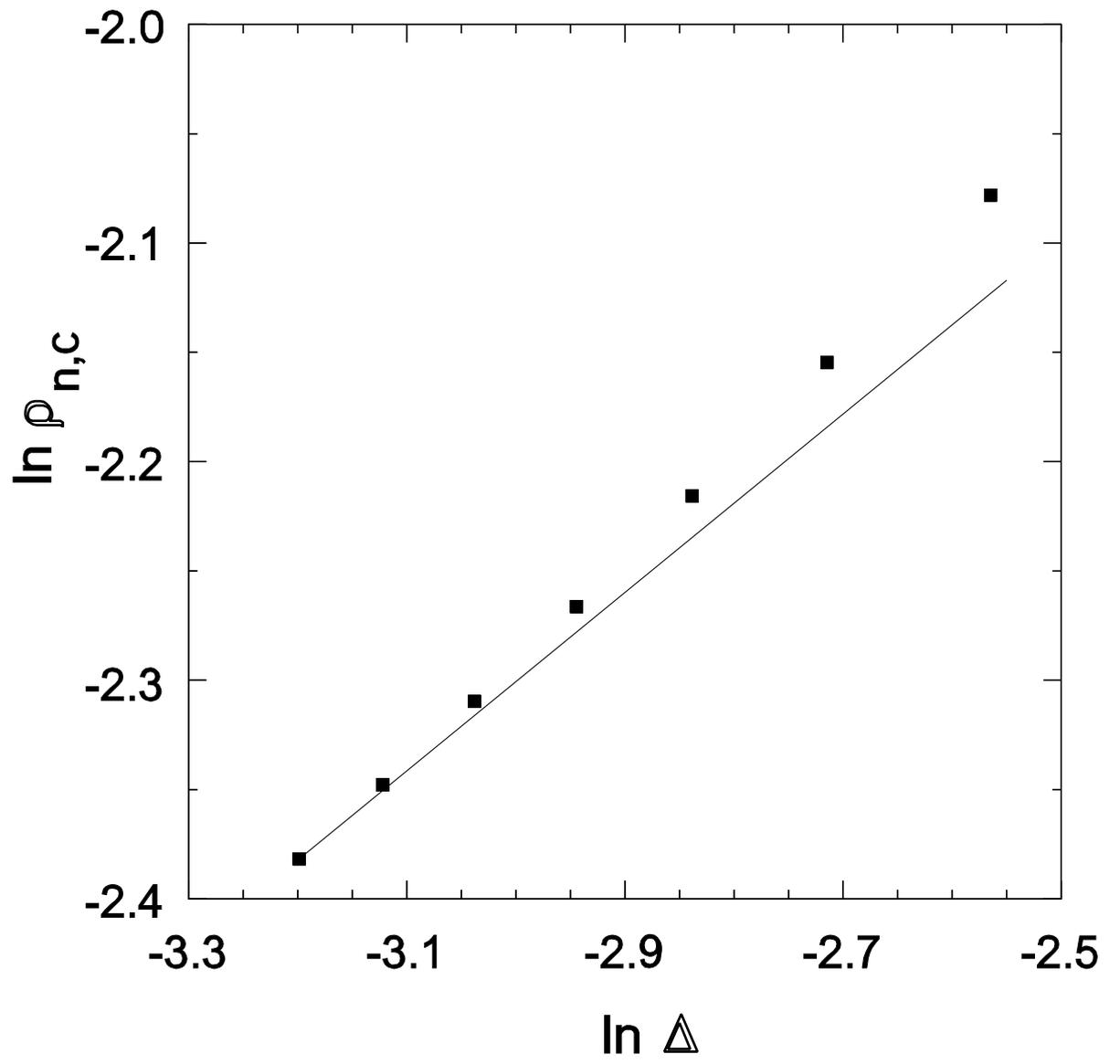
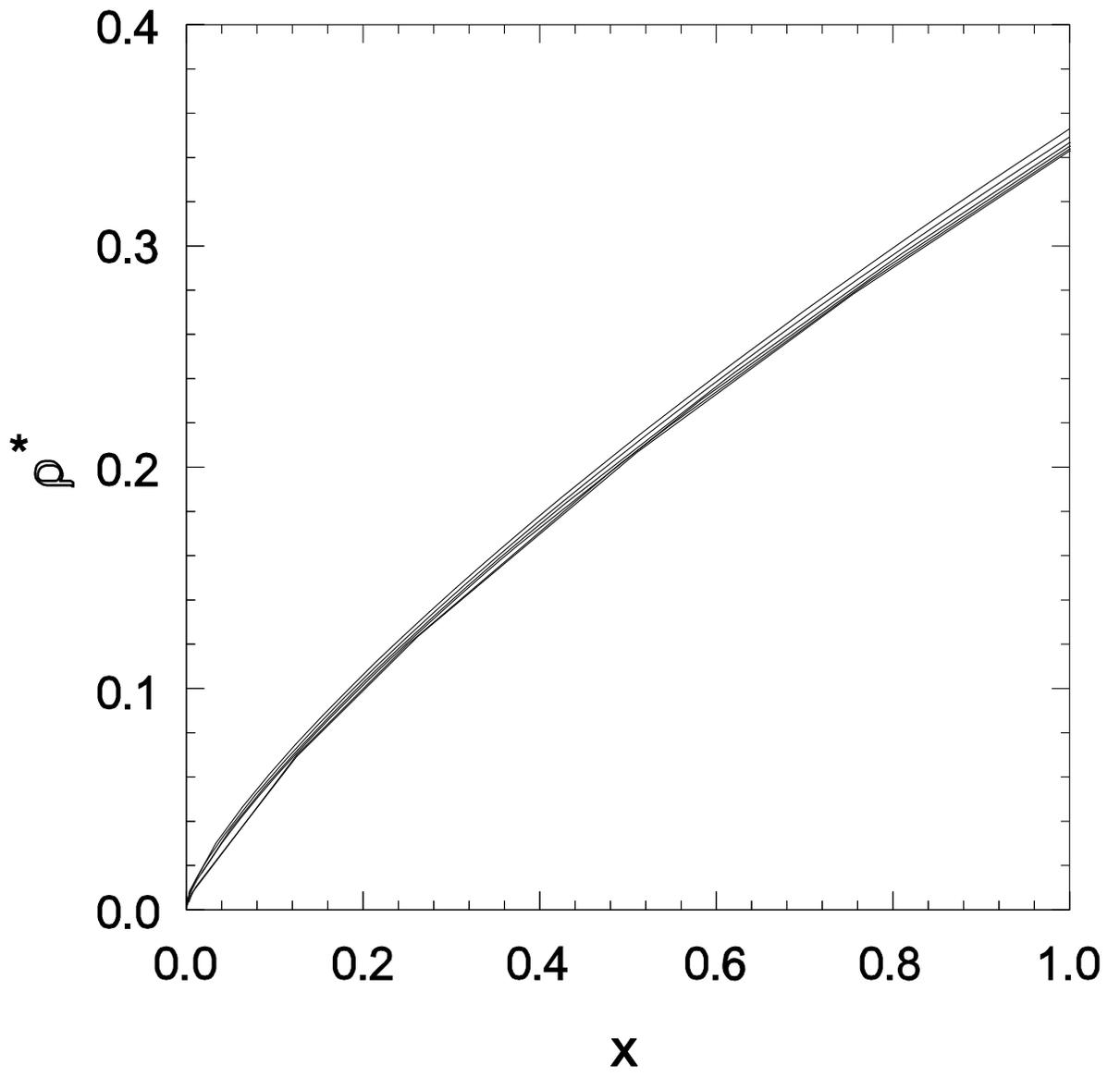
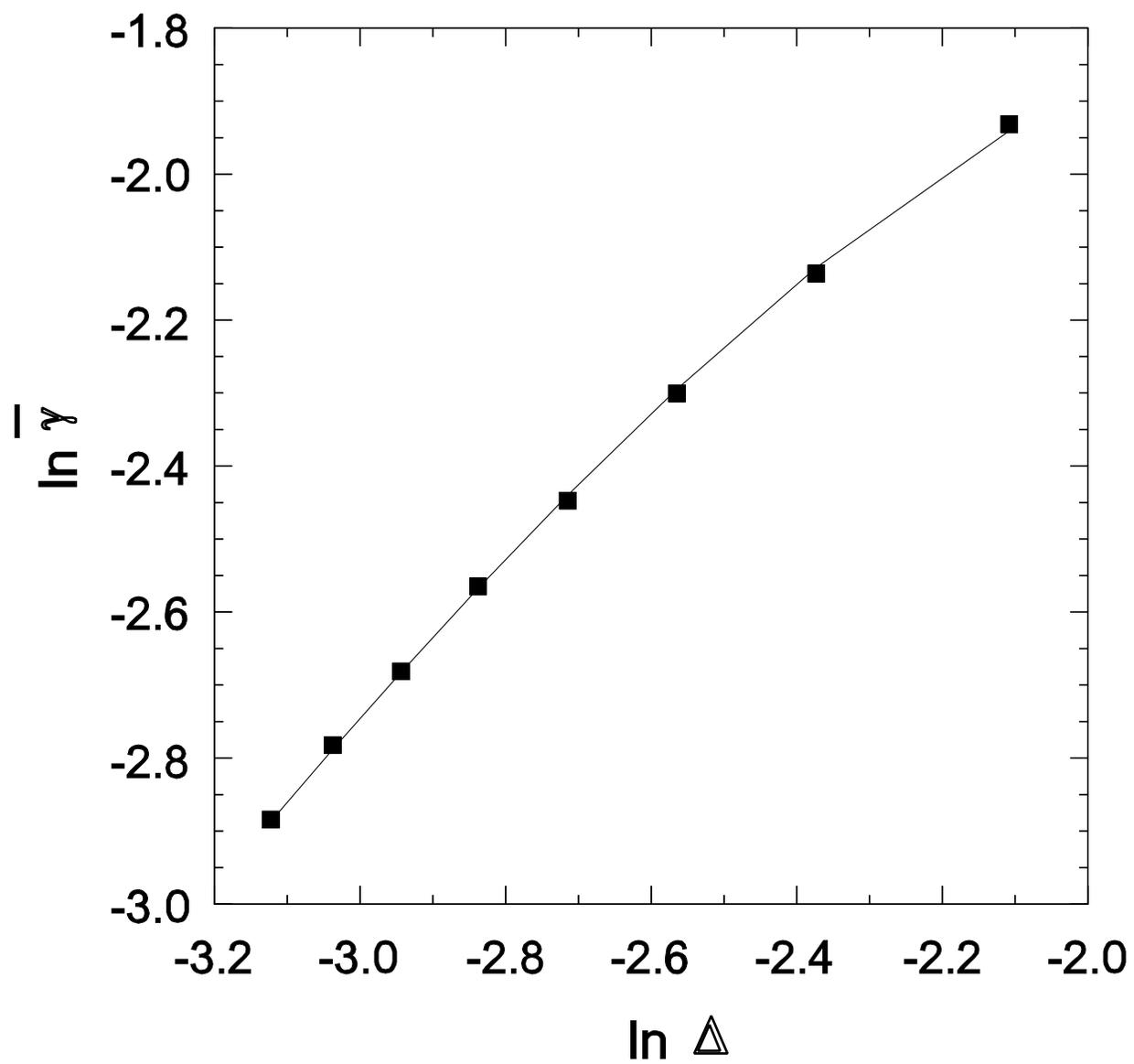
\end{document}